# 4D-UNet improves clutter rejection in human transcranial contrast enhanced ultrasound

Tristan Beruard, Armand Delbos, Arthur Chavignon, Maxence Reberol, Vincent Hingot

Resolve Stroke, Paris, France

**Abstract**

*Transcranial ultrasound imaging is limited by high skull absorption, limiting vascular imaging to only the largest vessels. Traditional clutter filters struggle with low signal-to-noise ratio (SNR) ultrasound datasets, where blood and tissue signals cannot be easily separated, even when the echogenicity of the blood is improved with contrast agents. Here, we present a novel 4D U-Net approach for clutter filtering in transcranial 3D Contrast Enhanced Ultrasound (CEUS) exploiting spatial and temporal information via a 4D-UNet implementation to enhance microbubble detection in transcranial data acquired in human adults. Our results show that the 4D-UNet improves temporal clutter filters. By integrating deep learning into CEUS, this study advances neurovascular imaging, offering improved clutter rejection and visualization. The findings underscore the potential of AI-driven approaches to enhance ultrasound-based medical imaging, paving the way for more accurate diagnostics and broader clinical applications.*

## 1 INTRODUCTION

Brain imaging with ultrasound has historically been constrained by the high absorption of ultrasound waves through the skull bone [1]. This challenge is particularly true when imaging slow blood flows and small vessels, which are masked by residual clutter caused by background signals originating from the skull and tissues. Moreover, modern microvascular imaging techniques increasingly use unfocused transmit beams to achieve higher frame rates, further reducing the signal-to-noise ratio (SNR). Even in contrast enhanced ultrasound (CEUS), where microbubbles (MB) enhance the echogenicity of blood, developing a good clutter filter remains a critical challenge to enable good microvascular imaging.

Developing an efficient clutter rejection filter to highlight the signal from blood is a decade old problem with no perfect solution [2]. A typical assumption is that stationary tissues contribute the most to this clutter whereas fast moving red blood cells compose the signal to extract. This way, the simplest method to perform clutter rejection is to apply a temporal filter in the temporal domain [3]. Yet, these filters work on short temporal sequences, limiting their ability to cut low frequencies and to exhibit the slowest flows. To address this issue, a new generation of filters based on matrix decomposition of blocs of frames using Singular Value Decomposition (SVD) or Principal Component Analysis (PCA) take advantage of longer ensemble lengths to improve the sensitivity to slower flows [4].

Although they are often called spatio-temporal clutter filters, they are not actually sensitive to spatial patterns within individual frames, unlike more recent algorithms based on machine learning. In particular, Convolutional Neural Networks (CNNs) can recognize spatial features in complex and noisy datasets and can further increase clutter filtering in CEUS [5] [6] [7]. In the context of CEUS,

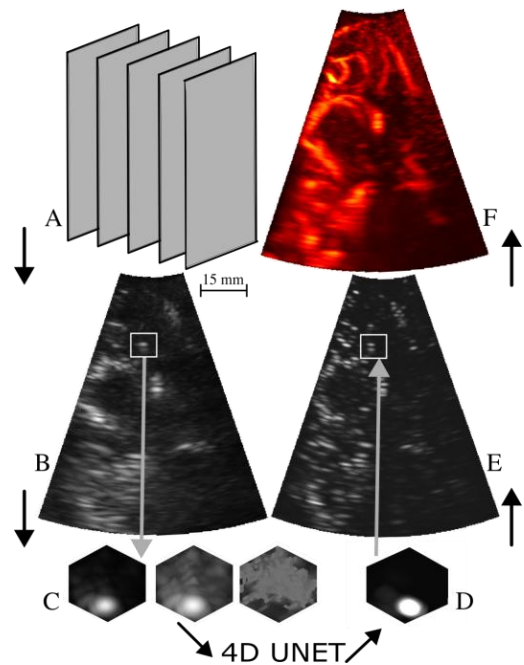

*Fig. 1: Complete decomposition of the clutter filtering pipeline, A. RF data, B. High-pass filter and beamforming, C. 5D slices, D. 3D+t slices, E. 3D+t complete volumes, F. Temporal accumulation*

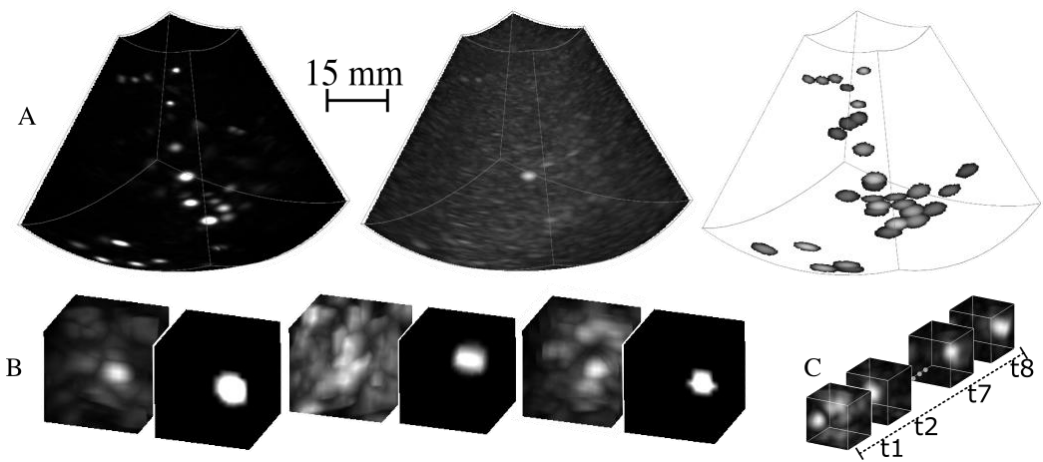

*Fig. 2: A. (from left to right): Beamformed MBs in water, beamformed addition of MBs with clutter, ground truth obtained from the MBs in water / B. Small patches with left: CF input of the model, right: Ground truth / C. Example of the trajectory of a MB in a patch*

CNNs can act as an additional contrast and resolution enhancer.

The input data for CNNs can be the raw radiofrequency (RF) data [8], [9], [10], [17] but most often follows a pre-processing step comprised of beamforming and clutter filtering. The output is generally a map displaying MB concentration or speed and representative of the blood volume or velocity [19].

Clutter filters based on CNNs rely on a training phase, in which labeled data is fed into the network. The labeled data originates from either simulated data [20] or the ground truth calculated using a regular clutter filter [21], meaning that those methods cannot be generalized for dataset where clutter filters are inefficient. In the human brain with unfocused transmit beams, clutter filters struggle to highlight individual MB and training methods are therefore not applicable.

In this study, we propose a new way to improve clutter filtering in transcranial 3D CEUS based on a training paradigm using real microbubbles signal in water as a training dataset. Additionally, we propose to use a 4D U-Net on 3D+t data to further exploit the spatial and temporal nature of the 3D CEUS data. This study is performed in a subset of transcranial data obtained in human patients with neurovascular alterations and an intact skull.

## 2   THEORY

In our implementation, schematized in Figure 1, the model's input is a block of preprocessed data. The input data is a 5D tensor where the three first dimensions are the 3D spatial volumes sampled by the probe and the fourth dimension is the evolution of these volumes through time. Finally, the last dimension corresponds to different "channels" used to train the model. The first channel contains the DAS output, the second channel is the coherence factor [22] while other channels are the high pass filtered L1 norm of the DAS, and its phase.

Then, this 5D tensor is cropped into smaller 5D volumes measured to fit the spatial expansion of a single MB through time. These small 5D cubes are used as the input of the model. The output is a 4D volume that is concatenated back to match the original volume.

### 2.1   Labelling strategy and training the model

A key limitation of existing implementations of CNN based clutter filtering is the generation of a labeled dataset to use as a ground truth for training. In some implementations, individual MB signals were used as ground truth for further training [13].

In our human clinical dataset, classical filters do not perform sufficiently well to detect microbubbles with a high contrast. Consequently, individual MB signal could not be used as a ground truth for supervised learning. Therefore, we propose a new training paradigm that can be adapted to any in vivo situation without needing a good contrast a priori.

To achieve this, we generate an artificial ground truth from in vitro data as represented in Figure 2. In this approach, two datasets, one of pure clutter, and another of pure MB signal are combined to create the training data. First, an acquisition of MBs in clear water is performed to generate a dataset with pure MB signal. Then, pure clutter signal is extracted from real in vivo datasets by selecting frames before MBs were injected. These two signals are then added together to form the composite dataset.

In this composite dataset, the position and intensity of the MBs can be easily computed and serve as ground truth. This way, a comprehensive pipeline is implemented to train the model with a wide number of MB + clutter configurations, with a loss function based on the input MB signal.

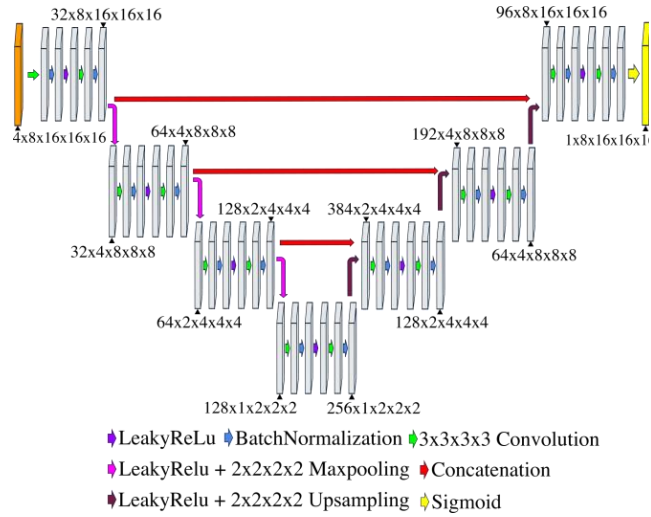

*Fig. 3: global architecture of the 4D U-Net*

The training data is highly specific to this use case, consequently there is no possibility to use a foundational model [23], or a pretrained base. The model is therefore trained from scratch. It is sufficiently small to be trained on a local GPU. Inspired by [24], we propose to train the network using small patches of the data, reducing compute-complexity and context-dependency.

## 2.2 Design of the CNN

This input/output choice imposes a regression architecture for the model. Other implementations emphasizing the temporal dimension showed promising results [25]. This way, the MBs' persistence and dynamics in the spatio-temporal domain can be exploited by the model.

Our implementation relies on a 4D U-Net which, to our knowledge, has not been used with CEUS data. We did not consider RNNs and Transformer based models [26], [27], [28], as the temporal information was sufficiently exploited with a 4-dimensional U-Net architecture. Additionally, the 4D-U-Net does not introduce any bias favoring the temporal dimension over spatial dimensions or vice versa, as its architecture treats spatial and temporal components in a structurally equivalent manner.

U-Nets are pertinent options for filtering because their encoder-decoder architecture effectively isolates background from the MBs while reconstructing a clean image [24],[29],[30]. More generally, comprehensive microvascular imaging implementations based on U-Nets have shown promising results in animal models, where clutter filtering performs better, due to higher frequency probes, low attenuation and controlled imaging conditions [31].

# 3 METHODS

The objective is to produce a deep learning pipeline to filter out the clutter and enhance microbubble signal in spatio-temporal samples on ultrasound acquisitions of the brain. There are three main points to focus on: acquiring a dataset for training, choosing the model architecture and training the model.

## 3.1 Data acquisition

### 3.1.1 Clinical data

The clinical data was gathered as part of the ESRIR study, reference NCT06793839 on clinicaltrials.gov. A set of 6 anonymized acquisitions were extracted from the ESRIR study database. Briefly, ultrasound acquisitions were performed at 1000 volumes per second on a 32x32 matrix with a central frequency of 2 MHz for 165s, representing a total of 165000 volumes. For each acquisition, a bolus of 2,4 ml of Sonovue was injected after 10s which means that the first 10000 frames are pure clutter without MB.

## A.2 Training data and Ground truth

Training data was created by combining datasets of MB signal and of clutter signal. Pure MB signal was acquired in a water tank lined with an absorbent at the bottom. MBs were diluted and stirred in the water tank until they visually appeared distinct enough on the images, typically separated by 1 cm. A gentle stir in the water generated a convection movement which randomized MB trajectories. 8 sets of 3200 frames were acquired with the same probe used to acquire the clinical data at a frame rate of 100 Hz. This clutter-free data was used to generate training inputs by adding to it pure clutter. The ground truth was obtained by cropping the easily detectable MBs.

Individual MBs were identified using the Trackpy library in Python [32]. The minmass parameter of the Trackpy detection was empirically set by using elbow detection on a plot representing the amount of object detected in different frames for multiple values of the minmass parameter. This method finds a compromise between not labelling noise as a MBs and detecting as many MBs as possible. The parameters of the detection function were a diameter of (5,5,5) and a separation of (7,7,7). The speeds of the bubbles were measured using Trackpy's linking function and were around 1 voxel per frame. The data was downsampled to achieve a more diverse distribution of MB speeds, with a uniform spread between 1 voxel/frame and 3 voxel/frame.

The linking function from Trackpy was then used again to reconstruct the trajectories of MBs across multiple frames. To focus the model on learning consistent trajectories over time, any MB trajectories shorter than 2 frames were removed from the ground truth.

To train the model, pure Radio Frequency (RF) MB signals ($X_{MBs}$) acquired in water were added to pure RF clutter signals ($X_{clutter}$) from the clinical study, prior to beamforming.

The relative intensity of the MB signal was modulated to form a uniform dataset using a linear coefficient $\lambda$. $X_{MBs}$ served to compute the ground truth. We assumed that the ultrasound medium is linear, meaning we neglected second order interactions between the clutter and the MBs.

### 3.2    Pre processing

#### 3.2.1    B.1 Beamforming

An initial step of clutter filtering was performed on the RF data by applying a simple high-pass filter, which removes from each frame the mean of a rolling window of 11 frames. The deep learning model will serve as a second clutter filtering step.

The data used as the input to the model was beamformed to obtain two channels: Coherence Factor (CF) and a complex-valued Delay And Sum (DAS). Classical CNN implementations do not accept complex values, consequently the DAS was split in three channels, one for its amplitude and two for the phase difference between the frame $t + 1$ and the frame $t$: $\Delta\Phi_{t+1} = \Phi_{t+1} - \Phi_t$. The dephasing is expressed in radians. To ensure continuity in the transition between $-2\pi$ and $2\pi$ and to avoid aliasing, it was split in two separated channels: $\cos(\Delta\Phi)$ and $\sin(\Delta\Phi)$. Overall, we ended up with four input channels, CF, DAS amplitude, dephasing cosine and dephasing sine.

#### 3.2.2    B.7 Cropping

We trained the model on small spatio-temporal windows (patches), to reduce computational cost and context dependency. The global output of the model is then given by the reconstruction using every small output window.

The size of the patch chosen was $(t, x, y, z) = (8, 16, 16, 16)$, which ensures that most bubble trajectories stay contained in each patch since their maximum speed is around 3 voxels/frame.

#### 3.2.3    B.8 Data processing

We performed data augmentation, by randomly swapping the spatial axes and randomly flipping the x and y axis. It helped generate more samples, with a greater variety of directions.

Standard normalization was applied to the CF and |DAS|. The Dephasing values were not normalized.

Normalization by dilation was applied to the ground truth, using a morphological dilation operation. This process consists in spatially expanding the values of the ground truth, to emphasize the MBs' maximum values. After dilation, each value of the original ground truth was divided by its corresponding dilated value. This normalization step enhanced the relative intensities and ensured that the intensity of the MB's center is set to 1.

Bubbles on the edges of the patches were removed from the ground truth so that the model would recognize complete MBs and ignore cropped MBs. The inference on real data is performed using an overlap between patches, which ensures that no MB will be ignored.

### 3.3    Model building

#### 3.3.1    C.1 Model architecture

We chose to use the U-Net architecture, introduced in [24], as the basis for the filtering network. The goal of the filter is to detect the MBs, which is achieved by separating them from the tissue. U-Nets have been shown to be very proficient in medical segmentation, making them an ideal candidate.

The model's architecture is composed of three main parts: the encoder, which extracts features from the input data; the bottleneck, which processes these features; and the decoder, which reconstructs a full-sized 4D output where bubbles are highlighted. Skip connections are used to concatenate outputs from different encoder levels to the corresponding decoder levels. These skip connections act as residual links, improving robustness to vanishing or exploding gradients [33] and helping to smooth the loss landscape [34].

The final activation function is a sigmoid from PyTorch's implementation since it is widely used in regression tasks and outputs the PSF of a MB normalized between 0 and 1.

#### 3.3.2    C.2 Implementation of 4D operators

The term 4D refers to the fact that the inputs are spatiotemporal. The decision to treat the temporal dimension the same way as the spatial dimensions stems from the intuition that, in 4D, microbubbles trajectories form tube-like structures, which can be more effectively localized by the network. 2D and 3D U-Nets are widely used in research and industrial settings. They can be directly implemented using PyTorch, Tensorflow or Jax. However, to our knowledge no library has implemented 4-dimensional operators for convolutional networks, they are limited to 2D or 3D data. This is why we built our own 4-dimensional operators to process the data.

For batch normalization, the 6D tensor $X$ of size $[B, C, L_x, L_y, L_z, T]$ (batch, channel, space, time) is reshaped in a 5D one of size $[B, C, L_x, L_y, L_z * T]$, which is passed through PyTorch's $BatchNorm3D$ operator. The normalized tensor can then be reshaped to its original size as the batch normalization operation preserves element order in PyTorch. With

$$BatchN_{4D}(X) = BatchN_{3D}(X.reshape(B, C, L_x, L_y, L_z * T)) \quad (1)$$

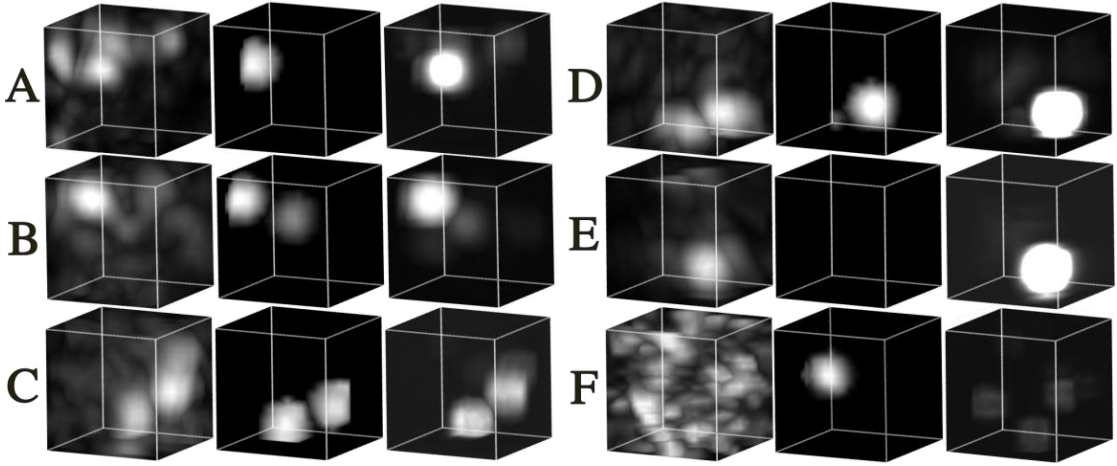

*Fig. 4: Examples of patches, Left to right: model input, ground truth, model output. A: example of the robustness of the model over spatial deformations on the input/ B and C: Detection of two MBs over one patch by the model / E: Example of a False Positive output /F: Example of a False negative, due to low MB intensity*

To upsample the tensor by factors $[i, j, k, l]$, we start by upsampling every 3D slice using PyTorch's built-in function with a scale factor of $[i, j, k]$. We then add $l - 1$ copies of the upsampled sub-tensor between the resulting 3D slices along the fourth dimension, effectively upsampling along this last dimension.

$$X_{upsampled} = [Upsample3D(X_{t=1}), Upsample3D(X_{t=1}), \dots, Upsample3D(X_{t=T})] \quad (2)$$

We tackle maxpooling by splitting the operation in two. For a pooling of kernel size $[i, j, k, l]$, each spatial 5D slice of the input is first passed through PyTorch's maxpool3d function with a kernel size of $[i, j, k]$. This results in a tensor of size $[b, c, \lfloor\frac{x}{i}\rfloor, \lfloor\frac{y}{j}\rfloor, \lfloor\frac{z}{k}\rfloor, t]$. We then apply the maxpool1d function along the time dimension for each spatial coordinate with a kernel size of $l$. This successfully creates the desired output array of size $[b, c, \lfloor\frac{x}{i}\rfloor, \lfloor\frac{y}{j}\rfloor, \lfloor\frac{z}{k}\rfloor, \lfloor\frac{t}{l}\rfloor]$.

$$X_{Maxpooled} = Pool1D([Pool3D(X_{t=1}), \dots, Pool3D(X_{t=T})]) \quad (3)$$

The implementation we used for convolutions was built and adapted from Jan Funke's git repository [33]. The 4D

convolution is a generalization of the 3D convolution. The complete formula with a 4D kernel size of $[i, j, k, l]$ and with $X'$ the resulting output is:

$$X'(b, c_{out}, x, y, z, t) = b_{c_{out}} + \sum_{c=1}^{C_{in}} \sum_{t_k, x_k, y_k, z_k} \omega_{j,c,t_k,x_k,y_k,z_k} * X_{b,c_{in},x+x_k,y+y_k,z+z_k,t+t_k} \quad (4)$$

With c' the output channel corresponding, j the position of $X$ in the batch, $C_{in}$ the number of input channels, and $(x, y, z, t)$ the positions of the output voxel corresponding to this formula. The repository's implementation uses 3D convolution, giving the following formula:

$$output_{t=i} = b + \sum_{t_k=-l//2}^{l//2} Conv_{3D}(X_{t=i+t_k}, \omega_{t=i}) \quad (5)$$

With b the full tensor of biases, $\omega$ the 6D full kernel tensor and output the 6D full output. The $Conv_{3D}$ operator from PyTorch operates over the channels and the batch dimensions in addition to the spatial one. The convolutions use a classical kernel size of $[3,3,3,3]$, stride 1 and padding 'same', both for the encoder and the decoder operations. MaxPooling and UpSampling operations use a window of size $[2,2,2,2]$. We used LeakyRelu as activation functions.

### 3.4 Model training

#### 3.4.1 D.1 Training parameters

The loss function chosen to train the model was the mean squared error, which is a classical choice for regression tasks. The learning rate of the Adam optimizer was initialized at $1.0 * 10^{-4}$, with a weight decay parameter of $1.0 * 10^{-4}$. The weights and biases were initialized using He initialization. The learning rate was carefully scheduled, by firstly setting it on constant $3.0 * 10^{-4}$, allowing the model to learn the global structure of the MBs trajectories, and then after X epochs when the Neural Network had sufficiently converged, it was decreased using the ReduceLROnPlateau scheduler to help the model learn the finer details of MB trajectories.

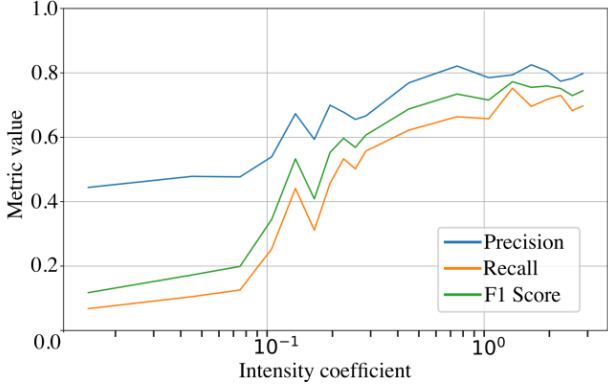

*Fig. 5: Scores of the 4D U-Net depending on the intensity coefficient that weights the MBs in the sum Tissue + MBs (log scale)*

### 3.4.2 D.2 Training summary

The model was trained around 100 epochs, with each epoch taking approximately 45 minutes on a Nvidia RTX 4080 with 16 Go of memory.

## 4 RESULTS

We present results divided into two categories: in vitro data, where the ground truth allows us to compute standard detection metrics; and in vivo human acquisitions, where the evaluation is qualitative.

### 4.1 Results over in vitro dataset

In the in vitro dataset, the ground truth is separable from the clutter which allows us to localize MBs. Therefore, it is possible to measure the performance of the model for detection on a test dataset. This test set is comprised of different acquisitions of MBs that were not used for training. The metrics used are calculated over the raw model output. To detect the MBs on the model's outputs, their centers were detected using a peak detection algorithm from SciPy, by setting an absolute threshold adjusted empirically, and then fixed for the whole test set (the peak intensity on every frame has to be greater than this threshold for a spot to be detected as a MB). The centers were detected frame by frame and not linked to form trajectories. This means that one MB trajectory contains multiple detected centers evolving through time. The metrics consider only the centers which are at least three voxels away from the border.

Comparing the positions of MBs on the model output to the ground truth, as seen for different situations on the Figure 4, allows us to calculate the amount of True Positive / False Positive / False Negative detections. The precision, the recall and the f1-score are computed using these metrics.

These scores are dependent on the intensity coefficient used to mix the raw MB data and the clutter signal. To quantify this dependence, the metrics are computed over different datasets with different values for this coefficient. Each test dataset is composed of 3200 samples (100 batches) extracted from data excluded from the training set.

As shown in Figure 5, the scores increase as the intensity coefficient increases, meaning the network detects MBs more accurately. This is logical as MBs are best detected when the ratio between their intensity and the intensity of clutter is high. At best, the metrics are capped between 0.7 and 0.8 and do not reach 1 because some MBs do not appear in the ground truth, and some ground truth annotations may not correspond to actual MBs. Additionally, even at high intensity, a few MBs remain barely visible in the training examples, which correspond to local areas with higher clutter noise.

The 4D U-Net exhibits limitations when MBs become poorly visible: the recall score, and consequently the f1-score, decrease significantly when the intensity coefficient becomes low. Typically, the threshold used for MB detection over clutter is in the order of -9 dB [34].

### 4.2 Results over in vivo dataset

The model is used on in vivo human data collected in a clinical trial. The complete acquisition volume is beamformed, obtaining all input channels. A high pass filtering operation is performed on the data before the beamforming and standardization is applied, exactly like on the training data.

The temporal volume is divided in 4D patches with a spatial overlap of 6 voxels between adjacent patches. To avoid a higher intensity on the border of output patches due to the summed overlap, voxel values on the borders are combined using Gaussian weights centered on each patch. The model's outputs for every patch are reassembled into a complete volume.

The high pass filtering + 4D U-Net method is compared to high pass filtered CF accumulation, and SVD filtered CF accumulation. The SVD is done by cutting off the 20 first eigenvalues of the temporal dimension over blocks of 256 frames. Figure 6 shows that the 4D-U-Net isolates MBs signal from the background better than SVD or high pass filtering.

Results are accumulated over 6400 acquisition frames, corresponding to 6,4s of acquisitions, along their corresponding frame indices. Specifically, for each frame position within a block, the resulting block contains the sum of all frames at the same position across the previous blocks.

Finally, the standard deviation is computed for each voxel in the resulting block, enhancing the visibility of MBs trajectories. These temporal accumulations can be seen on Figure 7. The vessels appear thinner in the model output, leading to better vessel separation compared to SVD and high pass filtering alone (Figure 7.C). Additionally, certain vascular structures are more clearly visible (Figure 7.C) in the model output than in the classical results, highlighting the model's high sensitivity.

## 5   DISCUSSION

The vascularization that is visible in the accumulations with high-pass filter and SVD filter can be improved by the model. This can be observed on the accumulations, and in some regions a better separation of the vessels is noticeable. A hypothesis to explain the behavior of the model is that the improvement possibly originates from the shape of the model's output, which contain smaller bubble than in the filtered data, which also have a wider and more uniform spherical shape. This would be a PSF thinning behavior.

Furthermore, certain vascular structures that are barely discernible after high-pass filtering appear more clearly in the model, indicating the model's higher sensitivity to certain MBs that are faintly visible in the noise by the model. This increase in sensitivity can be attributed to the model's improved recognition of characteristic spatial patterns of MBs and its robustness to temporal signal interruptions, which are common in in vivo data. The model enforces spatial patterns that are temporally propagated, like a MB moving, and can outperform temporal filters when MBs flicker or momentarily disappear. The individual spatial patterns of MBs are not considered in the SVD, which performs purely temporal filtering, albeit on projected singular spatial vectors, but without highlighting special spatial structures.

A limitation is that the model's accumulation appears more discontinuous than the accumulations of the high pass filtered, and not compounding the diffuse signal of undetected MBs which still contributes to the overall signal. This creates a sharper image, where only structures with strong MBs signal can be seen. This is advantageous for morphological measurements, but can be detrimental to perfusion or functional imaging, where

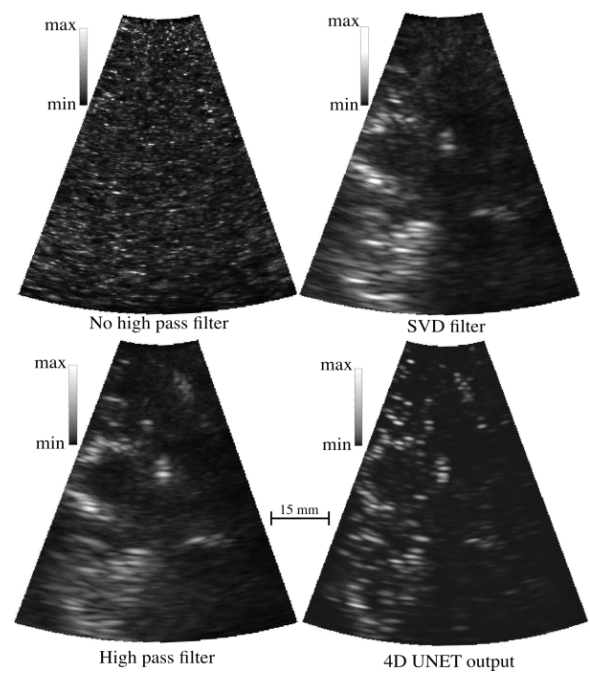

Fig. 6: Comparison of a temporal frame of 3D volumes (MIP view) between different filters steps. The three first volumes (in the order top left, top right, down left) are over the CF only the temporal changes of the background signal can be indicative of physiological changes.

Although all the data were acquired with the same probe and system, the deformations present in real in vivo data are not perfectly reproduced in the in vitro data used for training the model, as the MBs-tissue interactions were not modeled and thus neglected. The PSF of in vivo MB is therefore different from that of in vitro MBs, as the wavefront undergoes significant distortion during propagation through the tissue and while passing through the skull. This highlights a clear avenue for improving our method, which could involve the use of labeled in vivo data. One potential approach to enhance labeling could be leveraging the current 4D-UNet to provide an initial estimation of the MBs' positions, thereby facilitating manual labeling.

The model captures temporal coherence over a maximum of only 8 frames whereas SVD leverages long temporal windows to extract clutter characteristics. One potential improvement for the 4D-UNet model would be to increase the size of the temporal window to incorporate a broader temporal context.

Additionally, the model's input data is already filtered, which reduces the information available to the model compared to unfiltered data. Another possible enhancement would be to decrease the cutoff frequency of the high pass filter applied to the model's input, allowing it to exploit a larger information source. This partly explains why only MBs within large brain vasculature are visible, as they exhibit sufficiently fast dynamics to stay visible after the high-pass filter. Conversely, small brain vessels contain slow-moving MBs whose information is lost to the high-pass filter. Reducing the intensity of this filter while increasing the temporal window could allow the recovery of new trajectories within smaller vascular structures.

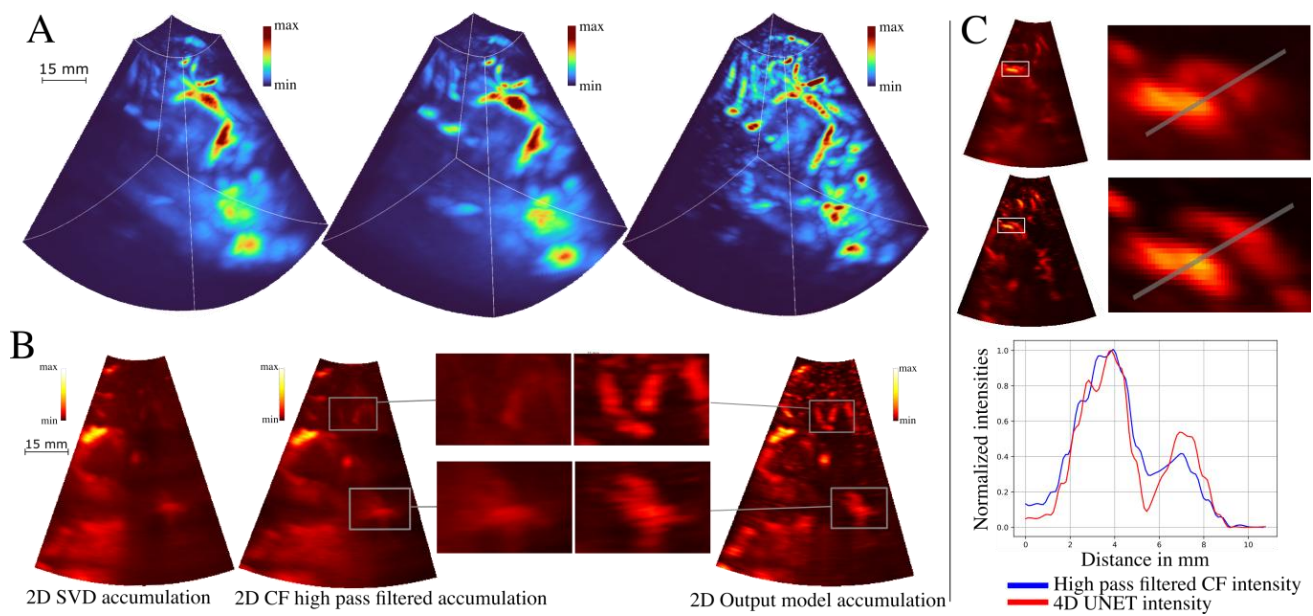

*Fig. 7: (A): 3D temporal accumulations, from Left to right: SVD filtered and CF beamformed data, High pass filtered and CF beamformed data, 4D UNET output over CF and DAS high pass filtered data. (B): 2D slice of temporal accumulation (C): Zoom over two vascularization and comparison of the spatial separation over the drawn segment between High pass filtered CF and 4D UNET output over CF and DAS high pass filtered data.*

Unfortunately, increasing the size of the temporal windows would require the use of different deep learning architectures, capable of handling long temporal dependencies, such as transformers.